\documentclass{article}
\usepackage{amsmath}
\usepackage{amssymb}
\usepackage{frascatiphys}
\usepackage{graphicx}
\begin{document}
\title{ 
THEORY OF OFF-AXIS AND GRAVITATIONAL WAVE EVENTS
}
\author{
H.J. van Eerten  \\
{\em Department of Physics, University of Bath, Claverton Down, Bath BA2 7AY, United Kingdom}
}
\maketitle
\baselineskip=11.6pt
\begin{abstract}
GW170817 was not merely an absolute breakthrough in gravitational wave astrophysics and a first in multi-messenger astronomy. The quality and diversity of the electro-magnetic counterpart emission is staggering on its own as well, including unprecedented kilonova spectra and a broadband off-axis gamma-ray burst afterglow that has progressed along a trajectory of rise and decay and by now has even been measured using very large baseline interferometry. For these proceedings, I will summarize the points for discussion that I presented during the workshop regarding off-axis short gamma-ray bursts and their (un-)successful jets and their emission. Given that developments are currently moving very fast in the field, I also touch on some results that have appeared in the literature following the Vulcano meeting.
\end{abstract}
\baselineskip=14pt

\section{Introduction}

The recent joint detection\cite{LIGO2017, Abbott2017} of electro-magnetic (EM) and gravitational wave (GW) emission from merging neutron stars (NS) has been one of the biggest scientific achievements of the past decades, revolutionising astrophysical observations and having far-ranging repercussions for transients high-energy astrophysics. Various predictions for a range of potentially detectable EM counterparts had been made prior to the detection of GW170817 / GRB 170817A, but the actual quality and quantity of the data wildly exceeded expectations: kilonovae had largely remained a theoretical construct up to this point, although the first tentative detections were getting published\cite{Tanvir2013,Berger2013}, while short gamma-ray bursts (sGRBs) and their subsequent afterglow jets were only expected for a subset of NS mergers due to their collimated nature. GRB 170817A of course provided us not just with a close view of a GW source, but also with a kilonova \emph{and} a GRB \emph{and} a long-lasting broadband afterglow. In these proceedings, I discuss in particular how the detection of the afterglow for this source has forced us to update our models for sGRB jets and afterglows.

\section{GW170817 / GRB 170817A prompt emission and gravitational waves}

Let us first recap the non-afterglow observations of GW170817 / GRB 170817A that nevertheless have an implication for our understanding of its afterglow. From the GW analysis, the angle $\theta_\nu$ between the angular momentum vector of the binary neutron has been inferred to be $\theta_\nu \approx 18^\circ \pm 8^\circ$ \cite{Abbott2017b, Mandel2018}, potentially pointing at an off-axis observation of the prompt and/or afterglow emission, in case the jet collimation angle $\theta_0 < \theta_\nu$. The source was detected in gamma rays, both by INTEGRAL\cite{Savchenko2017} and Fermi-GBM\cite{Goldstein2017}, before its gravitational wave signal was identified. The GRB was delayed by 1.7 s relative to the merger time. If the emission is indeed detected off-axis, this delay limits\cite{Abbott2017b, ShoemakerMurase2018, MatsumotoNakarPiran2018} the prompt emission radius $R_\gamma$ according to $t_{\gamma,\oplus} / (1+z) = t_\gamma - \cos(\theta_\nu - \theta_0) R_\gamma / c$, where $t_\gamma$ and $t_{\gamma,\oplus}$ the observed time in the burster and observer frame respectively, $z = 0.0098$ the redshift, $c$ the speed of light and $R_\gamma = \int_0^{t_\gamma} \beta_\gamma c \delta t$. For example, an offset $\theta_\nu - \theta_0 \approx 13^\circ$ would limit $R_\gamma \lesssim 2\times 10^{12}$ cm for $\beta_\gamma \uparrow 1$ (or smaller if the jet launching was delayed or the jet was moving non-relativistically before reaching $R_\gamma$). This radius is as expected\cite{Nakar2007} for internal shock dissipation, but at odds with Poynting flux dominated models or other models that put the dissipation closer to the deceleration radius.

While labeled `an ordinary short GRB'\cite{Goldstein2017}, GRB 170817A did exhibit some nonstandard features (one other example that comes close to its behavior is GRB 150101B\cite{Abbott2017b, Troja2018GRB150101B, Burns2018}). Its isotropic equivalent prompt energy release $E_{\gamma,iso} \approx 5 \times 10^{46}$ erg, rendering it 2 to 6 orders of magnitude less energetic than other sGRBs. The burst consisted of a brief ($\sim 0.5$ s) smooth pulse, followed by weak soft emission between 1 and 2 seconds after the GBM trigger time\cite{Goldstein2017, Veres2018}. The gamma-ray spectrum of the initial peak is well described by a power law turning over into an exponential drop (the `Comptonized function' among the standard fit functions typically used, rather than e.g. the `Band' function or a smoothly broken power law), peaking at $E_{peak} = 229\pm 78$ keV. It is noteworthy that this peak value is typical, whereas $E_{\gamma,iso}$ is not, making the burst an outlier in the $E_{peak} - E_{\gamma,iso}$ plane\cite{Ghirlanda2009}. The later emission is best fit by a blackbody spectrum peaking at $k_B T = 10.3 \pm 1.5$ keV. The smoothness of the initital pulse is consistent with an off-axis observation smoothing out variability relative to an on-axis observation\cite{Salafia2016}. The minimum variability timescale for the peak, (taken to be the rise time of $0.125$ s) yields an upper limit on the size of the emission region of $\delta R \sim 4 \times 10^{13}$ cm\cite{Goldstein2017}. 

The Lorentz factor $\Gamma$ at the time of emission can be constrained in a number of ways \cite{Kasliwal2017, Troja2017, MatsumotoNakarPiran2018, Veres2018}, mostly implying an emission site that is at least mildly relativistic. The low energy of GRB 170817A can be explained by positioning the observer outside of the cone of a collimated outflow, diminishing the observed flux through off-axis relativistic beaming. But the angular dependency of relativistic beaming is extremely steep, so much that a typical flow with $\Gamma \sim 100$ on-axis (as expected for short GRBs\cite{Nakar2007, Berger2014}), would lead to a predicted energy even lower than what is observed. This issue can be resolved by moving the observer closer to the axis of the jet, lowering the Lorentz factor of the outflow to open up the beaming cone, or both. The limiting cases of this are either placing the observer inside of the sub-energetic wings of a jetted flow, or having only mildly relativistic quasi-spherical flow. However, this is not straightforward either if one wishes to avoid opacity issues due to electron-positron pair-production and Thomson scattering\cite{Troja2017, LambKobayashi2018}), although not impossible\cite{Mooley2018superluminal}. It is therefore possible that the prompt emission observed for GRB 170817A has been produced by a mechanism different from ordinary short GRBs.

A discussion of the kilonova emission and properties lies beyond the scope of these notes. Nevertheless, the properties of the merger inferred from the kilonova emission also constrain the physics of the GRB prompt and afterglow. In particular, the are indications for a brief\cite{MetzgerThompsonQuataert2018} or extended\cite{YuLiuDai2018} period of energy injection by a magnetar-like remnant. For the latter case of a long-lasting remnant ($\gg 1$ second, which is not a natural fit given that the GW observations constrain internal magnetic fields to not be much lower than $\sim 10^{16}$ G\cite{Pooley2018}), short GRB afterglow plateaus come to mind, which have been argued to be produced by long-term energy injection from magnetar dipole spin-down. Any plateau stage for GRB 170817A would have been missed by an off-axis observer.

\section{GRB 170817A and jet / blast wave toy models}

Keeping the constraints from the non-afterglow measurements in mind, we now turn to the afterglow stage of GRB 170817A. The afterglow of GRB 170817A was first detected at radio and X-ray frequencies\cite{Hallinan2017, Troja2017} at around nine days, following a number of non-detections (upper limits) at these wavelengths\cite{Evans2017}. At the time when the first reports on this source became public, it was reasonable to interpret the data for these first two weeks as exhibiting signs of a flattening in the light curve, given that the flux levels for the second and third X-ray points were comparable after a rise relative to the first detection, and given that the radio detections were interspersed with another claimed upper limit\cite{Alexander2017}. The early non-detections made clear that we could not be witnessing an on-axis afterglow, whereas the apparent flattening was consistent with the peak emission from a basic collimated flow model. Here `basic' refers to the simplicity of the \emph{initial conditions}, a blast wave with an energy, mass and velocity distribution that is independent of angle up to some truncation angle $\theta_0$ (i.e. 'top-hat', the simplest non-spherical starting point). Off-axis observations with observers placed at $\theta_{obs} > \theta_0$ would see a steeply rising light curve as the relativistic beaming cone of the shock-accelerated electrons at the blast wave front come into view. Once the jet is into view completely, the light curve would turn over towards a decaying slope.

However, the explanation of the afterglow light curve observations as revealing the turnover of top-hat jet emission quickly turned out not to be viable when renewed observations showed a continuing rise with $\alpha \approx -0.8$ in $t^{-\alpha}$, lasting longer than the turnover would have. This compounded issues already identified with the prompt emission in view of off-axis top-hat jet models (i.e. the extremely steep angular dependence of relativistic beaming), that led authors to include structured jet alternatives in the first batch of papers\cite{Troja2017, Hallinan2017}. The latest observations\cite{Troja2018b}, indicate a rising slope of $\alpha \approx -0.9$, a peak time $t_{p} = 164\pm 12$ days and finally a steep decay slope $\alpha \approx 2$. To understand the options from here, a brief review of popular jet morphologies in the context of new observations is in order (for an extensive older review on jets, see \cite{Granot2007}).

{\bf \underline{basic top hat jets.}} The first jetted flow models\cite{Rhoads1997} for long and short GRBs simply assumed a forward shock moving ultra-relativistically and truncated at an angle that would start growing noticeably in the observer frame once the Lorentz factor dropped to $\gamma \approx 1 / \theta_0$ (in the observer frame, the sideways velocity is suppressed by this same factor $\gamma$; the turning point marks when spreading becomes clear to observers even when including this factor). Once spreading starts, this becomes a runaway effect due to the feedback between the increasing size of the working surface sweeping up external medium and the induced further decay of $\gamma$. For reasons of simplicity, this model includes no ininital angular jet structure (with radial structure either neglected in the case of homogeneous shell models, or taken from a self-similar solution\cite{BlandfordMcKee1976} before spreading). The evolution of such jets has been numerically simulated in detail, revealing the expected lateral structure of a shock system expanding sideways as well as forwards (see e.g. \cite{Meliani2007, ZhangMacFadyen2009, vanEerten2011, WygodaWaxmanFrail2011}). Fully numerically resolved simulations\cite{vanEertenMacFadyen2013} confirm that lateral expansion in these systems affects the decay phase slope of the light curve rather than the rising phase slope. For reasonable jet opening angles ($\theta_0 > 0.05$ rad), no runaway stage of expansion occurs in practice because the jet moves into the trans-relativistic stage approximately around the same time full causal contact across all angles is achieved (ultra-relativistic flow and full causal contact are necessary assumptions in runaway spreading models). In conclusion, off-axis top-hat jets remain ruled out by the rising slope, whether simulated numerically or solved (semi-) analytically.

{\bf \underline{Power-law structured jets.}} A straightforward extension of the top-hat model is to impose lateral structure on the jet using a discrete number of lateral components or simple functional form such as a power law dependency on angle for the energy, $\epsilon \propto \theta^{-a}$ (e.g. \cite{LipunovPostnovProkhorov2001, RossiLazzatiRees2002}). One of the initial motivations for such models was to shift the observed variation in jet opening angles and energetics to jet orientation relative to the observer instead\cite{LipunovPostnovProkhorov2001, ZhangMeszaros2002}. This requires that the lateral drop in energy in energy is not too steep ($a < 2$), but such a shallow slope also has the implication that the afterglow emission must be bright and decaying for off-axis observers still within the wings of the jet. Given the late-time rise of GRB 170817A, this possibility is ruled out as well, and the remaining options for the lateral jet structure are those where the wings have little energy relative to the tip, e.g. through a steep power-law or even exponential dependence of energy on angle.

{\bf \underline{Gaussian jets.}} Jets with a steep drop in energy $\epsilon$ with angle can also be modeled usings exponentials, i.e. $\epsilon \propto \exp [-\theta^2/2 \theta^2_c]$, with $\theta_c$ setting the width of the core. On-axis observers and observers at very small angles will observe a light curve that very closely resembles that produced by a top-hat jet (another practical reason why numerical simulations tended to start from top-hats), and observers at $\theta_{obs} \lesssim 4 \theta_c$ will see a declining signal from the beginning\cite{Rossi2004}. However, observers at larger angles will see a smoothly rising signal, as different regions come into view once the beaming cones of their relativistic emission have opened up sufficiently ---due to the deceleration of the jet--- to encompass the direction to the observer. During the rising stage this behaviour, where the emission is dominated by successively more energetic regions, is equivalent to observing a spherical blast wave with an ongoing injection of energy, and we return to this point below. Another attractive feature of (Gaussian) structured jets is that they manage to capture well the jet morphology produced by detailed numerical simulations of relativistic jet launching \cite{AloyJankaMueller2005} and breakout\cite{Morsony2007, MizutaAloy2009, Lazzati2017breakout, Kathirgamaraju2018, Margutti2018}. Once the entire jet has come into view, the effect will be similar to the top-hat jet case, and a steep decay of the light curve is seen comparable to that of a GRB observed on-axis post jet-break\cite{Rhoads1999, SariPiranHalpern1999}.

{\bf \underline{Quasi-spherical blast waves with energy injection.}} A spherical blast wave model for the afterglow using regular afterglow parameters would show a bright decelerating light curve already at early times. On the other hand, the gamma-rays and afterglow of GRB 170817A might have been produced by a special subclass of GRB, or perhaps by something not normally detected as a GRB at all. One option here would be the (quasi-)spherical release of energy with only moderate velocity, perhaps a cocoon of material (e.g. neutron star merger debris\cite{Nagakura2014}) which has managed to absorb the energy of what would otherwise have been a collimated GRB jet. Initial predictions for quasi-spherical cocoon models included afterglow peak timescales of days\cite{NakarPiran2017, Lazzati2017}, rather than the observed weeks for GRB 170817A. This timescale is set by the deceleration radius of the ejecta. If the peak time is to match that observed for GRB 170817A, the deceleration time needs to be artificially postponed. A natural means to do so, would again be to invoke additional structure in the outflow, this time radial rather than lateral. Indeed, the right peak times can be achieved by assuming a stratification of cocoon outflow velocities behind the front of the shock, effectively acting as a mechanism for the long-term injection of energy\cite{Mooley2018}. There is ample precedent for models including velocity stratification, both for GRBs\cite{ReesMeszaros1998} and kilonova ejecta\cite{Metzger2017}. During the rising light curve stage, it is fundamentally not possible to distinguish observationally between radial energy injection and lateral energy `injection' (i.e. more energetic regions coming into view). However, following the peak, the light curve slope is not expected to be steep\cite{Troja2018}, and to fall well short of the recently measured value\cite{Troja2018b}.

 Of all jet toy models, it turns out that the Gaussian jet plus off-axis observer configuration is the most successful in capturing the observed features of GRB 170817A\cite{Troja2018, Troja2018b}, from initial non-detections to extended shallow rise and turnover to steep decay. Basic cocoon models would have peaked on a timescale comparable to the kilonova. Relativistic jets with extended energetic wings would have been visible too soon. The top-hat jet profile is unable to account for an extended shallow rise. Quasi-spherical (cocoon) models with radial energy injection will ultimately end up decaying according to the non-relativistic Sedov-Taylor solution for a non-relativistic point explosion\cite{FrailWaxmanKulkarni2000}, perhaps slightly steeper on account of lateral spreading, when merely wide-angled rather than quasi-spherical\cite{LambMandelResmi2018}. Either slope will fall short of the observed slope by a large margin, and will only be achieved at larger timescales than currently observed anyway.
 
\subsection{Successful jets, choked jets and cocoons}

While a staple of jet modeling of active galactic nuclei and long GRBs for decades, cocoons are a new arrival when it comes to the modeling of short GRBs\cite{Murguia-Berthier2014, Nagakura2014}. Cocoons are produced when (part of) the jet energy gets dissipated in a dense medium before emergence into the more dilute circumburst environment. It is an open question whether enough dense material, for example debris from the neutron star merger, is present in the path of the jet to materially effect its evolution. Numerical studies of the ejection of merger debris show that this can be concentrated along the orbital plane\cite{Rosswog1999}, but also more isotropic in the case of a soft neutron star matter equation of state\cite{Hotokezaka2013}. The jet can be choked completely if the combination of jet opening angle and merger debris is sufficiently high (e.g. total quasi-isotropic ejecta mass of $10^{-2} M_\odot$, combined with a wide-angled jet injection of $~45^\circ$, depending on jet power\cite{Nagakura2014}). In the choked jet case, a quasi-spherical trans-relativistic blast wave is formed.

As stated above, a choked jet scenario appears ruled out by recent observations, but this does not imply the absence of a cocoon entirely. Successful jet breakout with a prominent cocoon component has been simulated by various groups\cite{Kasliwal2017, GottliebNakarPiran2018, Lazzati2018, Xie2018}. This leads to a structured jet that can be modeled using the semi-analytical approaches already described. When it comes to the broadband afterglow light curve of GRB 170817A, the data is by this point consistent with a successfully launched jet with significant angular structure that potentially includes a cocoon component.

\section{Synchrotron emission from the afterglow}

\begin{figure}[htb]
    \begin{center}
        {\includegraphics[scale=0.6]{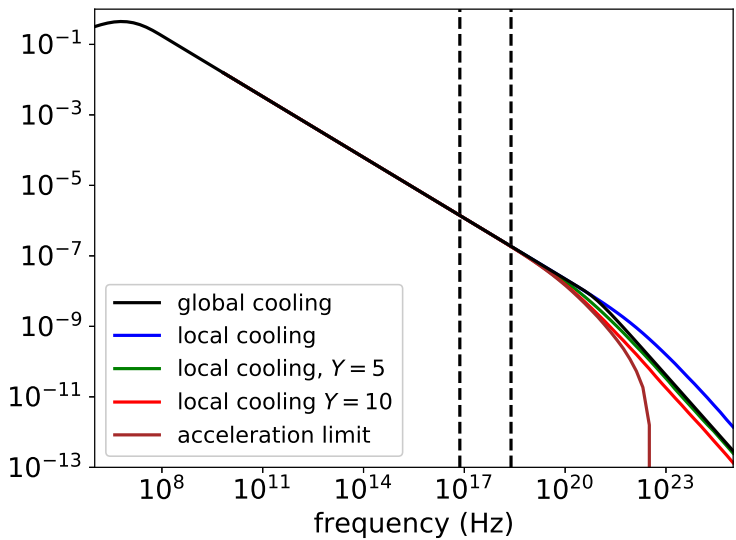}}\hspace{0.5cm}
        \caption{Synthetic spectra for an off-axis structured jet GRB with physics settings comparable to those inferred for GRB 170817A around 150 days, using different approximations to electron cooling. Here $Y$ is the Compton $Y$-parameter, to account for an additional impact on the electron cooling rate from synchrotron self-Compton losses.}
\label{spectrum_figure}
    \end{center}
\end{figure}

There is little doubt that the predominant emission mechanism during the afterglow phase is synchrotron emission. Afterglow detections covering nine orders of magnitude in frequency (i.e. from radio to X-rays) are fully consistent with a single power law spectrum\cite{Troja2017, Haggard2017, Troja2018, Margutti2018, Alexander2018, Ruan2018, Troja2018b}. Interpreted as being part of the synchrotron spectrum between the \emph{injection break} $\nu_m$ (associated with the lower cut-off of the shock-accelerated power-law electron population) and the \emph{cooling break} $\nu_c$ (beyond which the impact of the synchrotron energy loss term on the electron population becomes noticeable), the measured spectral slope $\beta = 0.585 \pm0.005$ implies an electron power law slope $-p = -2.17 \pm 0.01$. The nine orders of magnitude represent a remarkable stretch for a single power law: radio observations (albeit on-axis ones) often fall below $\nu_m$ initially, and the cooling break $\nu_c$ is often found between optical and X-rays for afterglows\cite{Greiner2011}. Nevertheless, once the off-axis orientation of the observer is taken into account, the afterglow spectra for GRB 170817A can be fit using otherwise reasonable values for the model parameters\cite{Troja2017, LambKobayashi2018, Troja2018, Troja2018b}. Figure \ref{spectrum_figure} illustrates the broader synchrotron spectrum for variables in this range, including turnover points. A cooling break this high could actually be less relevant to the spectrum than the upper cut-off on the emission resulting from the balance between particle acceleration timescale and synchrotron loss timescale for electrons at very high energies (i.e. the electrons predominantly responsible for high frequency photon emission). Approximations of this timescale are typically based on a comparison between electron gyroradius and characteristic time for synchrotron energy loss\cite{DaiCheng2001}. Two further complications to modeling electron cooling are that there can also be additional cooling due to synchrotron self-Compton losses, and that estimates for the cooling break are highly sensitive to the level of detail that is applied to modeling the \emph{local} cooling rate of electrons advected from the shock front\cite{GranotSari2002, vanEertenZhangMacFadyen2010, Guidorzi2014}.

The value $p = 2.17$ lies in between the two typical limiting values for $p$ expected from theoretical considerations. For particle shock-acceleration in non-relativistic blast waves, $p = 2$ is expected\cite{Bell1978, BlandfordOstriker1978, BlandfordEichler1987}. For the case of ultra-relativistic blast waves, various authors have argued for $p \approx 2.2-2.4$, based on particle-in-cell simulations and (semi-)analytical methods\cite{Kirk2000, Achterberg2001, Spitkovsky2008}. It is therefore tempting to interpret\cite{Margutti2018} the precise measurement of $p = 2.17 \pm 0.01$ for GRB 170817A directly within this framework as being indicative of a moderately relativistic blast wave Lorentz factor with a Lorentz factor $\Gamma \approx 3-10$. However, decades of afterglow modeling (as well as modeling of other synchrotron sources such as blazars), have established an overall observational range and distribution of $p$ values that is at minimum observationally at odds with this direct interpretation. Multiple authors have demonstrated that the distribution of $p$ as measured for various samples of afterglows, is not consistent with a universal value\cite{ShenKumarRobinson2006, Curran2009, Curran2010}, nor does $p$ exhibit a general hardening trend over time when spectral energy distributions are compared over multiple epochs\cite{Varela2016} (such a trend which would manifest itself both in a changing spectral slope and in the light curve temporal slope). As a matter of fact, $p$ has even been measured\cite{DaiCheng2001, PanaitescuKumar2001, Filgas2011, ZhangHuangZong2015} to have values $p < 2$ including measurements using broadband observations covering many orders in magnitude (e.g. $p = 1.73 \pm 0.03$\cite{Varela2016}). It is not (yet) clear whether the view of particle acceleration allowed by an off-axis event such as GRB 170817A is fundamentally more clear than that for the larger sample of shock-accelerating systems.

\section{Closing remarks}

In all, the long-term evolution of the broadband afterglow light curve of GRB 170817A has been remarkably successful. The late rise was instrumental in ruling out an observer positioned within the opening angle of a relativistic outflow, implying either an observer positioned outside of the jet, or no successful jet and a trans-relativistic quasi-spherical explosion. The continued rise then ruled out off-axis jet models with hard edges and single-shell cocoon models, requiring a structured jet or velocity stratification within a cocoon model. Ultimately, the turn-over to steep decay served to rule-out quasi-spherical models entirely.

In addition, we now even have very large baseline interferometry (VLBI) observations for the same source. The apparent source size at 207.4 days has been constrained to be less than 2 milliarcseconds\cite{Ghirlanda2018}, which is too small to be consistent with an isotropic, mildly relativistic blast wave such as the cocoon-type models described above. Other VLBI observations\cite{Mooley2018superluminal} reveal superluminal motion consistent with the energetic and narrow core of a structured jet. Taken together, the afterglow evidence is becoming compelling that, like other short GRBs, GRB 170817A came with a collimated relativistic jet, and that the GW170817 / GRB 170817A event has become the smoking gun for the short GRB / neutron star merger connection. Observations of future event will tell whether there is on occasion sufficient material in the path of the outflow to give rise to sub-population of choked jet explosions.

However, even if the afterglow can be reconciled with the `typical' expected behaviour for short GRBs, the prompt emission does remain puzzling and atypical. Observationally, the energy is very low, but neither off-axis beaming nor low Lorentz factor explanations are a completely satisfactory solution. It remains an open question whether the prompt emission has been produced by the same mechanism as that observed for short GRBs in general. And even in the case the prompt emission was produced differently, it is not clear whether the normal process genuinely did not occur or was merely not detectable at our observational angle.


\begin{thebibliography}{99}
\bibitem{LIGO2017}LIGO Scientific Collaboration and Virgo Collaboration, Phys. Rev. Lett. {\bf 119}, 161101 (2017).
\bibitem{Abbott2017}B.P.~Abbott {\it et al}, ApJL {\bf 848}, L12 (2017).
\bibitem{Abbott2017b}B.P.~Abbott {\it et al}, ApJL {\bf 848}, L13 (2017).
\bibitem{Achterberg2001}A.~Achterberg {\it et al}, MNRAS {\bf 328}, 393 (2001)..
\bibitem{Alexander2017}K.D.~Alexander {\it et al}, GCN circular 21851 (2017).
\bibitem{Alexander2018}K.D.~Alexander {\it et al}, ApJL {\bf 863}, L18 (2018).
\bibitem{AloyJankaMueller2005}M.A.~ Aloy, H.T~Janka and E.~M{\"u}ller, {A\&A} {\bf 436}, 273 (2005).
\bibitem{Berger2013}E.~Berger, W.~Fong and R.~Chornock, ApJL {\bf 774}, L23, (2013).
\bibitem{Berger2014}E.~Berger, ARoAA {\bf 52}, 43 (2014).
\bibitem{Bell1978}A.R.~Bell, MNRAS {\bf 182}, 147 (1978).
\bibitem{BlandfordOstriker1978}R.D.~Blandford and J.P.~Ostriker, ApJL {\bf 221}, L29 (1978).
\bibitem{BlandfordEichler1987}R.~Blandford and D.~Eichler, PhR {\bf 154}, 1 (1987).
\bibitem{BlandfordMcKee1976}R.D.~Blandford and C.F.~McKee, Physics of Fluids {\bf 19}, 1130 (1976).
\bibitem{Burns2018}E.~Burns {\it et al}, arXiv:1807.02866v1 (2018).
\bibitem{Curran2009}P.A.~Curran {\it et al}, MNRAS {\bf 395}, 580 (2009).
\bibitem{Curran2010}P.A.~Curran {\it et al}, ApJL {\bf 716}, L135 (2010).
\bibitem{DaiCheng2001}Z.G.~Dai and K.S.~Cheng, ApJL {\bf 558}, L109 (2001).
\bibitem{Evans2017}P.A.~Evans {\it et al}, Science {\bf 358}, 1565 (2017)
\bibitem{Filgas2011}R.~Filgas {\it et al}, A\&A {\bf 535}, A57 (2011).
\bibitem{FrailWaxmanKulkarni2000}D.A.~Frail, E.~Waxman and S.R.~Kulkarni, ApJ {\bf 537}, 191 (2000).
\bibitem{Ghirlanda2009}G.~Ghirlanda {\it et al}, A\&A {\bf 496}, 585 (2009).
\bibitem{Ghirlanda2018}G.~Ghirlanda {\it et al}, arXiv:1808.00469 (2018).
\bibitem{Goldstein2017}A.~Goldstein {\it et al}, ApJL {\bf 848}, L14 (2017).
\bibitem{GottliebNakarPiran2018}O.~Gottlieb, E.~Nakar and T.~Piran, MNRAS {\bf 473}, 576 (2018).
\bibitem{Greiner2011}J.~Greiner {\it et al}, A\&A {\bf 526}, A30 (2011).
\bibitem{GranotSari2002}J.~Granot and R.~Sari, ApJ {\bf 568}, 820 (2002).
\bibitem{Granot2007}J.~Granot, RMxAC {\bf 27}, 140 (2007).
\bibitem{Guidorzi2014}C.~Guidorzi {\it et al.}, MNRAS {\bf 438}, 752 (2014).
\bibitem{Haggard2017}D.~Haggard {\it et al.}, ApJL {\bf 848}, L25 (2017).
\bibitem{Hallinan2017}G.~Hallinan {\it et al.}, Science {\bf 358}, 1579 (2017).
\bibitem{Hotokezaka2013}K.~Hotokezaka {\it et al}, PHRvd {\bf 87}, 024001 (2013).
\bibitem{Kasliwal2017}M.M.~Kasliwal {\it et al}, Science {\bf 358}, 6370 (2017).
\bibitem{Kathirgamaraju2018}A.~Kathirgamaraju, R.~Barniol Duran and D.~Giannios, MNRAS {\bf 473}, L121
\bibitem{Kirk2000}J.G.~Kirk {\it et al.}, ApJ {\bf 542}, 235 (2000).
\bibitem{LambKobayashi2018}G.P.~Lamb and S.~Kobayashi, MNRAS {\bf 478}, 733 (2018).
\bibitem{LambMandelResmi2018}G.P.~Lamb, I.~Mandel and L.~Resmi, arXiv:1806.03843 (2018).
\bibitem{Lazzati2017}D.~Lazzati {\it etl al.}, MNRAS {\bf 471}, 1652 (2017).
\bibitem{Lazzati2017breakout}D.~Lazzati {\it et al.}, ApJL {\bf 848}, L6 (2017).
\bibitem{Lazzati2018}D.~Lazzati {\it et al.}, PhRvL {\bf 120}, 241103 (2018).
\bibitem{LipunovPostnovProkhorov2001}V.M.~Lipunov, K.A.~Postnov and M.E.~Prokhorov, Astron. Rep. {\bf 45}, 236 (2001).
\bibitem{Mandel2018}I.~Mandel, ApjL {\bf 853}, L12 (2018).
\bibitem{Margutti2018}R.~Margutti {\it et al.}, ApJL {\bf 856}, L18 (2018).
\bibitem{MatsumotoNakarPiran2018}T.~Matsumoto, E.~Nakar and T.~Piran, ArXiv:1807.04756v1 (2018).
\bibitem{Meliani2007}Z.~Meliani {\it et al.}, MNRAS {\bf 376}, 1189 (2007).
\bibitem{Metzger2017}B.~Metzger, Living Reviews in Relativity {\bf 20}, 3 (2017).
\bibitem{MetzgerThompsonQuataert2018}B.D.~Metzger, T.A.~Thompson and E.~Quataert, ApJ {\bf 856}, 101 (2018).
\bibitem{MizutaAloy2009}.A~Mizuta and M.A.~Aloy, ApJ {\bf 699}, 1261 (2009).
\bibitem{Mooley2018}K.P.~Mooley {\it et al.}, Nature {\bf 554}, 207 (2018).
\bibitem{Mooley2018superluminal}K.P.~Mooley {\it et al.}, arXiv:1806.09693 (2018).
\bibitem{Morsony2007}B.J.~Morsony, D.~Lazzati and M.C.~Begelman, ApJ {\bf 665}, 569 (2007).
\bibitem{Murguia-Berthier2014}A.~Murguia-Berthier {\it et al.}, ApJ {\bf 788}, L8 (2014).
\bibitem{Nagakura2014}H.~Nagakura {\it et al.}, ApJL {\bf 784}, L28 (2014).
\bibitem{Nakar2007}E.~Nakar, Phys Rep {\bf 442}, 166 (2007).
\bibitem{NakarPiran2017}E.~Nakar and T.~Piran, ApJ {\bf 834}, 28 (2017).
\bibitem{PanaitescuKumar2001}A.~Panaitescu and P.~Kumar, ApJL {\bf 560}, L49 (2001).
\bibitem{Pooley2018}D.~Pooley {\it et al.}, ApJL {\bf 859}, L23 (2018).
\bibitem{ReesMeszaros1998}M.~Rees and P.~M\'esz\'aros, ApJ {\bf 496}, L1 (1998).
\bibitem{Rhoads1997}J.E.~Rhoads, ApJ {\bf 487}, L1 (1997).
\bibitem{Rhoads1999}J.E.~Rhoads, ApJ {\bf 525}, 737 (1999).
\bibitem{RossiLazzatiRees2002}E.~Rossi, D. Lazzati and M.J.~Rees, MNRAS {\bf 332}, 945 (2002).
\bibitem{Rossi2004}E.~Rossi {\it et al.}, MNRAS {\bf 354}, 86 (2004).
\bibitem{Rosswog1999}S.~Rosswog {\it et al.}, A\&A {\bf 341}, 499 (1999).
\bibitem{Ruan2018}J.J~Ruan {\it et al.}, ApJL {\bf 853}, L4 (2018).
\bibitem{Salafia2016}O.S.~Salafia {\it et al}, MNRAS {\bf 461}, 3607 (2016).
\bibitem{SariPiranHalpern1999}R.~Sari, T.~Piran and J.~Halpern, ApJ {\bf 519}, L17 (1999).
\bibitem{Savchenko2017}V.~Savchenko {\it et al}, ApJL {\bf 848}, L15 (2017).
\bibitem{ShenKumarRobinson2006}R.~Shen, P.~Kumar and E.L.~Robinson, MNRAS {\bf 371}, 1441 (2006).
\bibitem{ShoemakerMurase2018}I.M.~Shoemaker and K.~Murase, Phys Rev D {\bf 97}, 083013 (2018).
\bibitem{Spitkovsky2008}A.~Spitkovsky, ApJL {\bf 682}, L5 (2008).
\bibitem{Tanvir2013}N.R.~Tanvir {\it et al}, Nature {\bf 500}, 547 (2013).
\bibitem{Troja2017}E.~Troja {\it et al}, Nature {\bf 551}, 71 (2017).
\bibitem{Troja2018}E.~Troja {\it et al}, MNRAS Letters {\bf 478}, L18 (2018).
\bibitem{Troja2018b}E.~Troja {\it et al}, arXiv:1808.06617 (2018).
\bibitem{Troja2018GRB150101B}E.~Troja {\it et al}, arXiv:1806.10624 (2018).
\bibitem{vanEertenZhangMacFadyen2010}H.~van Eerten, W.~Zhang and A.~MacFadyen, ApJ {\bf 722}, 235 (2010).
\bibitem{vanEerten2011}H.~van Eerten {\it et al.}, MNRAS {\bf 410}, 2016 (2011).
\bibitem{vanEertenMacFadyen2013}H.~van Eerten and A.~MacFadyen, ApJ {\bf 767}, 141 (2013).
\bibitem{Varela2016}K.~Varela {\it et al}, A\&A {\bf 589}, A37 (2016).
\bibitem{Veres2018}P.~Veres {\it et al}, ArXiv:1802.07328v1 (2018).
\bibitem{WygodaWaxmanFrail2011}N.~Wygoda, E.~Waxman and D.~Frail, ApJ {\bf 738}, L23 (2011).
\bibitem{Xie2018}X.~Xie, J.~Zrake and A.~MacFadyen, ApJ {\bf 863}, 58 (2018).
\bibitem{YuLiuDai2018}Y.W.~Yu, L.D.~Liu and Z.G.~Dai, ApJ {\bf 861}, 114 (2018).
\bibitem{ZhangHuangZong2015}Q.~Zhang, Y.F.~Huang and H.S.~Zong, ApJ {\bf 811}, 83 (2015).
\bibitem{ZhangMacFadyen2009}W.~Zhang and A.~MacFadyen, ApJ {\bf 698}, 1262 (2009).
\bibitem{ZhangMeszaros2002}B.~Zhang, P.~M\'esz\'aros, ApJ {\bf 571},  876 (2002).
\end{thebibliography}
\end{document}